\documentclass[ALICE,manyauthors]{cernphprep}
\usepackage[comma,square,numbers,sort&compress]{natbib}
\usepackage{hyperref}
\usepackage{lineno}
\usepackage{xspace}
\usepackage[utf8]{inputenc}
\usepackage{siunitx}
\usepackage{verbatim}

\usepackage[T1]{fontenc}

\begin{document}
%

\newcommand{\pp}           {pp\xspace}
\newcommand{\ppbar}        {\mbox{$\mathrm {p\overline{p}}$}\xspace}
\newcommand{\XeXe}         {\mbox{Xe--Xe}\xspace}
\newcommand{\PbPb}         {\mbox{Pb--Pb}\xspace}
\newcommand{\pPb}          {\mbox{p--Pb}\xspace}
\newcommand{\AuAu}         {\mbox{Au--Au}\xspace}
\newcommand{\dAu}          {\mbox{d--Au}\xspace}

\newcommand{\snn}          {\ensuremath{\sqrt{s_{\mathrm{NN}}}}\xspace}
\newcommand{\pt}           {\ensuremath{p_{\rm T}}\xspace}
\newcommand{\meanpt}       {$\langle p_{\mathrm{T}}\rangle$\xspace}
\newcommand{\ycms}         {\ensuremath{y_{\rm CMS}}\xspace}
\newcommand{\ylab}         {\ensuremath{y_{\rm lab}}\xspace}
\newcommand{\etarange}[1]  {\mbox{$\left | \eta \right |~<~#1$}}
\newcommand{\yrange}[1]    {\mbox{$\left | y \right |~<~#1$}}
\newcommand{\dndy}         {\ensuremath{\mathrm{d}N_\mathrm{ch}/\mathrm{d}y}\xspace}
\newcommand{\dndeta}       {\ensuremath{\mathrm{d}N_\mathrm{ch}/\mathrm{d}\eta}\xspace}
\newcommand{\avdndeta}     {\ensuremath{\langle\dndeta\rangle}\xspace}
\newcommand{\dNdy}         {\ensuremath{\mathrm{d}N_\mathrm{ch}/\mathrm{d}y}\xspace}
\newcommand{\Npart}        {\ensuremath{N_\mathrm{part}}\xspace}
\newcommand{\Ncoll}        {\ensuremath{N_\mathrm{coll}}\xspace}
\newcommand{\dEdx}         {\ensuremath{\textrm{d}E/\textrm{d}x}\xspace}
\newcommand{\RpPb}         {\ensuremath{R_{\rm pPb}}\xspace}

\newcommand{\nineH}        {$\sqrt{s}~=~0.9$~Te\kern-.1emV\xspace}
\newcommand{\seven}        {$\sqrt{s}~=~7$~Te\kern-.1emV\xspace}
\newcommand{\twoH}         {$\sqrt{s}~=~0.2$~Te\kern-.1emV\xspace}
\newcommand{\twosevensix}  {$\sqrt{s}~=~2.76$~Te\kern-.1emV\xspace}
\newcommand{\five}         {$\sqrt{s}~=~5.02$~Te\kern-.1emV\xspace}
\newcommand{\twosevensixnn}{$\sqrt{s_{\mathrm{NN}}}~=~2.76$~Te\kern-.1emV\xspace}
\newcommand{\fivenn}       {$\sqrt{s_{\mathrm{NN}}}~=~5.02$~Te\kern-.1emV\xspace}
\newcommand{\onethirtynn}       {$\sqrt{s_{\mathrm{NN}}}~=~130$~Ge\kern-.1emV\xspace}
\newcommand{\twohundrednn}       {$\sqrt{s_{\mathrm{NN}}}~=~200$~Ge\kern-.1emV\xspace}
\newcommand{\seventev}        {$\sqrt{s}~=~7$~Te\kern-.1emV\xspace}

\newcommand{\LT}           {L{\'e}vy-Tsallis\xspace}
\newcommand{\GeVc}         {\ensuremath{\mathrm{GeV}/c}\xspace}
\newcommand{\MeVc}         {Me\kern-.1emV/$c$\xspace}
\newcommand{\GeVmass}      {Ge\kern-.2emV/$c^2$\xspace}
\newcommand{\MeVmass}      {Me\kern-.2emV/$c^2$\xspace}
\newcommand{\lumi}         {\ensuremath{\mathcal{L}}\xspace}
\newcommand{\geantf}       {\ensuremath{\textsc{Geant4}}\xspace}

\newcommand{\ITS}          {\rm{ITS}\xspace}
\newcommand{\TOF}          {\rm{TOF}\xspace}
\newcommand{\ZDC}          {\rm{ZDC}\xspace}
\newcommand{\ZDCs}         {\rm{ZDCs}\xspace}
\newcommand{\ZNA}          {\rm{ZNA}\xspace}
\newcommand{\ZNC}          {\rm{ZNC}\xspace}
\newcommand{\SPD}          {\rm{SPD}\xspace}
\newcommand{\SDD}          {\rm{SDD}\xspace}
\newcommand{\SSD}          {\rm{SSD}\xspace}
\newcommand{\TPC}          {\rm{TPC}\xspace}
\newcommand{\TRD}          {\rm{TRD}\xspace}
\newcommand{\VZERO}        {\rm{V0}\xspace}
\newcommand{\VZEROA}       {\rm{V0A}\xspace}
\newcommand{\VZEROC}       {\rm{V0C}\xspace}
\newcommand{\Vdecay} 	   {\ensuremath{V^{0}}\xspace}

\newcommand{\ee}           {\ensuremath{e^{+}e^{-}}} 
\newcommand{\pip}          {\ensuremath{\pi^{+}}\xspace}
\newcommand{\pim}          {\ensuremath{\pi^{-}}\xspace}
\newcommand{\kap}          {\ensuremath{\rm{K}^{+}}\xspace}
\newcommand{\kam}          {\ensuremath{\rm{K}^{-}}\xspace}
\newcommand{\pbar}         {\ensuremath{\rm\overline{p}}\xspace}
\newcommand{\kzero}        {\ensuremath{{\rm K}^{0}_{\rm{S}}}\xspace}
\newcommand{\lmb}          {\ensuremath{\Lambda}\xspace}
\newcommand{\almb}         {\ensuremath{\overline{\Lambda}}\xspace}
\newcommand{\Om}           {\ensuremath{\Omega^-}\xspace}
\newcommand{\Mo}           {\ensuremath{\overline{\Omega}^+}\xspace}
\newcommand{\X}            {\ensuremath{\Xi^-}\xspace}
\newcommand{\Ix}           {\ensuremath{\overline{\Xi}^+}\xspace}
\newcommand{\Xis}          {\ensuremath{\Xi^{\pm}}\xspace}
\newcommand{\Oms}          {\ensuremath{\Omega^{\pm}}\xspace}

\newcommand{\SigmaInelAntiD}{\ensuremath{\sigma_{\rm inel}(\overline{\rm d})}\xspace}
\newcommand{\SigmaInelAntiP}{\ensuremath{\sigma_{\rm inel}(\overline{\rm p})}\xspace}
\newcommand{\SigmaInelD}{\ensuremath{\sigma_{\rm inel}(\rm d)}\xspace}
\newcommand{\SigmaInelP}{\ensuremath{\sigma_{\rm inel}(\rm p)}\xspace}
\newcommand{\RatioAntid}{\ensuremath{\overline{\rm d}/{\rm d}}\xspace}
\newcommand{\RatioAntip}{\ensuremath{\overline{\rm p}/{\rm p}}\xspace}
\newcommand{\Antip}{\ensuremath{\overline{\rm p}}\xspace}
\newcommand{\Antid}{\ensuremath{\overline{\rm d}}\xspace}
\newcommand{\AntiHeThree}{\ensuremath{^{3}\overline{\rm He}}\xspace}
\newcommand{\HeThree}{\ensuremath{^{3}{\rm He}}\xspace}
\newcommand{\AntiHeFour}{\ensuremath{^{4}\overline{\rm He}}\xspace}
\newcommand{\HeFour}{\ensuremath{^{4}{\rm He}}\xspace}

\newcommand{\eg}{e.\,g.\xspace}
\newcommand{\ie}{i.\,e.\xspace}

\newcommand{\mb}{\ensuremath{~\si{\milli\barn}}\xspace}
\newcommand{\mub}{\ensuremath{~\si{\micro\barn}}\xspace}

\newcommand{\mbinv} {\ensuremath{~\si{\per\milli\barn}}\xspace}
\newcommand{\mubinv} {\ensuremath{~\si{\per\micro\barn}}\xspace}
\newcommand{\nbinv} {\ensuremath{~\si{\per\nano\barn}}\xspace}
\newcommand{\pbinv} {\ensuremath{~\si{\per\pico\barn}}\xspace}
\newcommand{\fbinv} {\ensuremath{~\si{\per\femto\barn}}\xspace}
\newcommand{\Lumi}{\ensuremath{\mathcal{L}}\xspace}

\begin{titlepage}
\PHyear{2020}       
\PHnumber{078}      
\PHdate{14 May}  

\title{Measurement of the low-energy antideuteron inelastic cross section}   

\Collaboration{ALICE Collaboration\thanks{See Appendix~\ref{app:collab} for the list of collaboration members}}
\ShortAuthor{ALICE Collaboration} 

\begin{abstract}

In this Letter, we report the first measurement of the inelastic cross section for antideuteron-nucleus interactions at low particle momenta, covering a range of $0.3 \leq p < 4$ \GeVc.
The measurement is carried out using \pPb collisions at a center-of-mass energy per nucleon--nucleon pair of \fivenn, recorded with the ALICE detector at the CERN LHC and utilizing the detector material as an absorber for antideuterons and antiprotons.
The extracted raw primary antiparticle-to-particle ratios are compared to the results from detailed ALICE simulations based on the \geantf toolkit for the propagation of \mbox{(anti)}particles through the detector material. The analysis of the raw primary \mbox{(anti)}proton spectra serves as a benchmark for this study, since their hadronic interaction cross sections are well constrained experimentally.
The first measurement of the inelastic cross section for antideuteron--nucleus interactions averaged over the ALICE detector material with atomic mass numbers $\langle A \rangle$ = 17.4 and 31.8 is obtained. The measured inelastic cross section points to a possible excess with respect to the Glauber model parameterization used in \geantf in the lowest momentum interval of $0.3 \leq p < 0.47$ \GeVc up to a factor 2.1.
This result is relevant for the understanding of antimatter propagation and the contributions
to antinuclei production from cosmic ray interactions within the interstellar medium. In addition, the momentum range covered by this measurement is of particular importance to evaluate signal predictions for  indirect dark-matter searches.

\end{abstract}
\end{titlepage}

\setcounter{page}{2} 



\setlength{\emergencystretch}{8pt}

The possible presence of antinuclei in the Milky Way could be explained either by reactions of \mbox{high-energy} cosmic rays with the interstellar medium or by more exotic sources, such as dark-matter annihilation~\cite{Cirelli:2014qia}.
Some dark-matter models \cite{Baer:2005tw,Donato:2008yx,Bertone:2010zza,Ibarra:2012cc,PhysRevD.62.043003} predict that low-energy antideuterons are a promising probe for indirect dark-matter searches since the contributions from cosmic-ray interactions in the energy range below 1-2 GeV per nucleon \cite{Duperray:2005si,Fornengo:2013osa,Aramaki:2015pii} are expected to be rather small.
For this reason, the search for antinuclei has been intensified in recent years with new satellite and balloon-borne experiments such as AMS-02~\cite{Kounine:2012ega} and GAPS~\cite{Hailey:2009fpa}.
So far, only antiprotons have been detected in space \cite{Aguilar:2016kjl}
and no clear evidence of heavier antinuclei production has been found yet~\cite{Fuke:2005it,Abe:2012tz}, but dedicated analyses searching for antideuteron and antihelium are currently ongoing~\cite{Donato:2008yx,Korsmeier:2017xzj}.

In order to get a reliable baseline for antideuteron production at low energies, realistic models of \mbox{cosmic-ray} transport are necessary.
In addition, also the predicted flux of antinuclei from dark-matter annihilation depends on the production mechanism and antinuclei transport properties within the interstellar medium.
There are three main relevant mechanisms that determine the signal and background rates: i) the antideuteron production, either in p--A and A--A reactions between cosmic rays and the interstellar medium depending on the element abundance, or in dark-matter annihilation processes,
ii) the antideuteron propagation in the galaxy, the heliosphere and the Earth’s atmosphere and iii) inelastic processes such as nuclear breakup, charge exchange or annihilation that occur during propagation and in experiments inside the detectors.
These three mechanisms must be measured as precisely as possible to interpret correctly any future measurement in satellite and balloon-borne experiments.
While the propagation has been constrained by measuring different nuclei from primary and secondary cosmic rays~\cite{Moskalenko:1997gh,Strong:2007nh,Boschini:2017fxq,Boschini:2018baj}, accelerator experiments can be used to study the production and the inelastic scattering cross sections.

Antimatter is copiously produced in high-energy collisions of protons and heavy ions \cite{Armstrong:2000gd,Anticic:2011ny}. This environment is hence well suited to study antinuclei properties.
At RHIC, the STAR and PHENIX Collaborations have measured \Antip, \Antid, \AntiHeThree and \AntiHeFour~\cite{Agakishiev:2011ib, Adam:2019wnb, Adler:2004uy,Adler:2001uy} yields employing \AuAu collisions at center-of-mass energies per nucleon-nucleon pair  of \onethirtynn and \twohundrednn.
At the LHC, the ALICE Collaboration has studied \Antip, \Antid, \AntiHeThree and \AntiHeFour production in \pp, \pPb and \PbPb collisions at center-of-mass energies per nucleon pair from 0.9 to 13 TeV~\cite{Adam:2015vda,Adam:2015pna,Acharya:2017fvb,Acharya:2017bso,Acharya:2019rgc,Acharya:2019rys,Acharya:2020sfy} and the yields obtained for $A\geq 2$ have been interpreted by means of coalescence or statistical hadronization models~\cite{Schwarzschild:1963zz,Kapusta:1980zz,Sun:2018mqq,Bellini:2018epz}. 
The LHC measurements combined with different coalescence models have been employed to estimate the antideuteron and antihelium flux from cosmic-ray interactions measurable by the AMS-02 and GAPS experiments~\cite{Kachelriess:2019taq,Kachelriess:2020uoh,Blum:2017qnn,Korsmeier:2017xzj}.
Since the inelastic cross sections for antinuclei--nuclei interactions are  measured precisely only for \Antip but barely known for heavier antinuclei, all the available calculations rely on poorly constrained parameterizations. 
For antideuterons, the inelastic cross sections have been measured for several materials only for two momentum values, $p = 13.3$ \GeVc~\cite{Denisov:1971im} and $p = 25$ \GeVc~\cite{Binon:1970yu}. However, the \mbox{low-momentum} range accessible by ALICE ($p\leq 5$ \GeVc) remains unexplored. For antihelium, no measurement of inelastic cross sections is available.

In this Letter we present a method to evaluate the inelastic cross section of antinuclei based on the measurement of raw reconstructed antiparticle-to-particle ratios.
Using ratios instead of individual particle yields allows to extract the antideuteron and antiproton cross sections independently from their production cross sections and for a broad momentum range.
We report the first measurement of the inelastic cross section for antideuteron--nucleus interactions in the momentum range of $0.3 \leq p < 4$ \GeVc.
The results presented are based on data collected during the 2016 \pPb LHC run at \fivenn. The performance of the ALICE detector and the description of its subsystems can be found in~\cite{Aamodt:2008zz,Abelev:2014ffa}.
Collision events are selected by using the information from the V0 detector, which consists of two plastic scintillator arrays located on both sides of the interaction point at forward and backward pseudorapidities.
A simultaneous signal in both arrays was used as a minimum-bias (MB) trigger.
In total, about 600 million MB events are selected for further analysis, which correspond to an integrated luminosity of $\Lumi_{\rm{int}}^{\rm{MB}} = 287\mubinv$, with a relative uncertainty of 3.7\%~\cite{Abelev:2014epa}.

The charged-particle tracks are reconstructed in the ALICE central barrel with the Inner Tracking System (ITS) and the Time Projection Chamber (TPC), which are located within a solenoid that provides a homogeneous magnetic field of $\SI{0.5}{\tesla}$ in the direction of the beam axis.
The ITS consists of six cylindrical layers of silicon detectors located at radial distances from the beam axis between $\SI{3.9}{\centi\meter}$ and $\SI{43}{\centi\meter}$.
The TPC extends radially from $r=\SI{85}{\centi\meter}$ to $r=\SI{247}{\centi\meter}$, is $\SI{5}{\meter}$ long and was filled with an Ar-CO$_{2}$ gas mixture during the 2016 data taking period. 
These two subsystems provide full azimuthal coverage for charged-particle trajectories in the pseudorapidity range $|\eta_{\rm lab}| < 0.8$. 
The selected tracks must fulfill basic quality criteria established in antinuclei analyses in p--Pb collisions~\cite{Acharya:2019rys}.
These criteria guarantee a resolution of about $2\%$ on the momentum reconstructed at the primary vertex ($p_{\mathrm{primary}}$) in this analysis.

The TPC is also used for the particle identification (PID) of \mbox{(anti)}protons and \mbox{(anti)}deuterons via their specific energy loss \dEdx in the gas volume, with a resolution of about $5\%$~\cite{Alme:2010ke}.
The $n(\sigma^{\rm TPC}_{i})$ variable represents the PID response in the TPC expressed in terms of the deviation between the measured and expected \dEdx for a particle species $i$, normalized by the detector resolution $\sigma$. The expected \dEdx is computed with a parameterized Bethe-Bloch curve \cite{Abelev:2014ffa}.
\mbox{(Anti)}protons and \mbox{(anti)}deuterons are selected by applying the selection criterion $|n(\sigma^{\rm TPC}_{i})| < 3$.
This selection is sufficient to obtain a purity close to 100\% for \mbox{(anti)}protons and \mbox{(anti)}deuterons in the momentum range below $0.7$ \GeVc and $1.4 $ \GeVc, respectively.
For the momentum range above $0.7$ \GeVc for \mbox{(anti)}protons and $0.9$ \GeVc for \mbox{(anti)}deuterons, the PID is complemented by the Time-Of-Flight (TOF) system, consisting of Multi-Gap Resistive Plate Chambers.
\mbox{(Anti)}proton and \mbox{(anti)}deuteron candidates selected in the TPC are matched to TOF hits and fits to the squared-mass distributions are performed for different momentum intervals~\cite{Acharya:2019rys}.
The PID purity in all momentum intervals is found to be higher than 88\% and 47\%
for the \mbox{(anti)}proton and \mbox{(anti)}deuteron samples, respectively.
The background is subtracted from the squared-mass spectra with a two-component fit~\cite{Acharya:2019rys}.

\begin{figure}[t]
\centering
\includegraphics[width=0.49\linewidth]{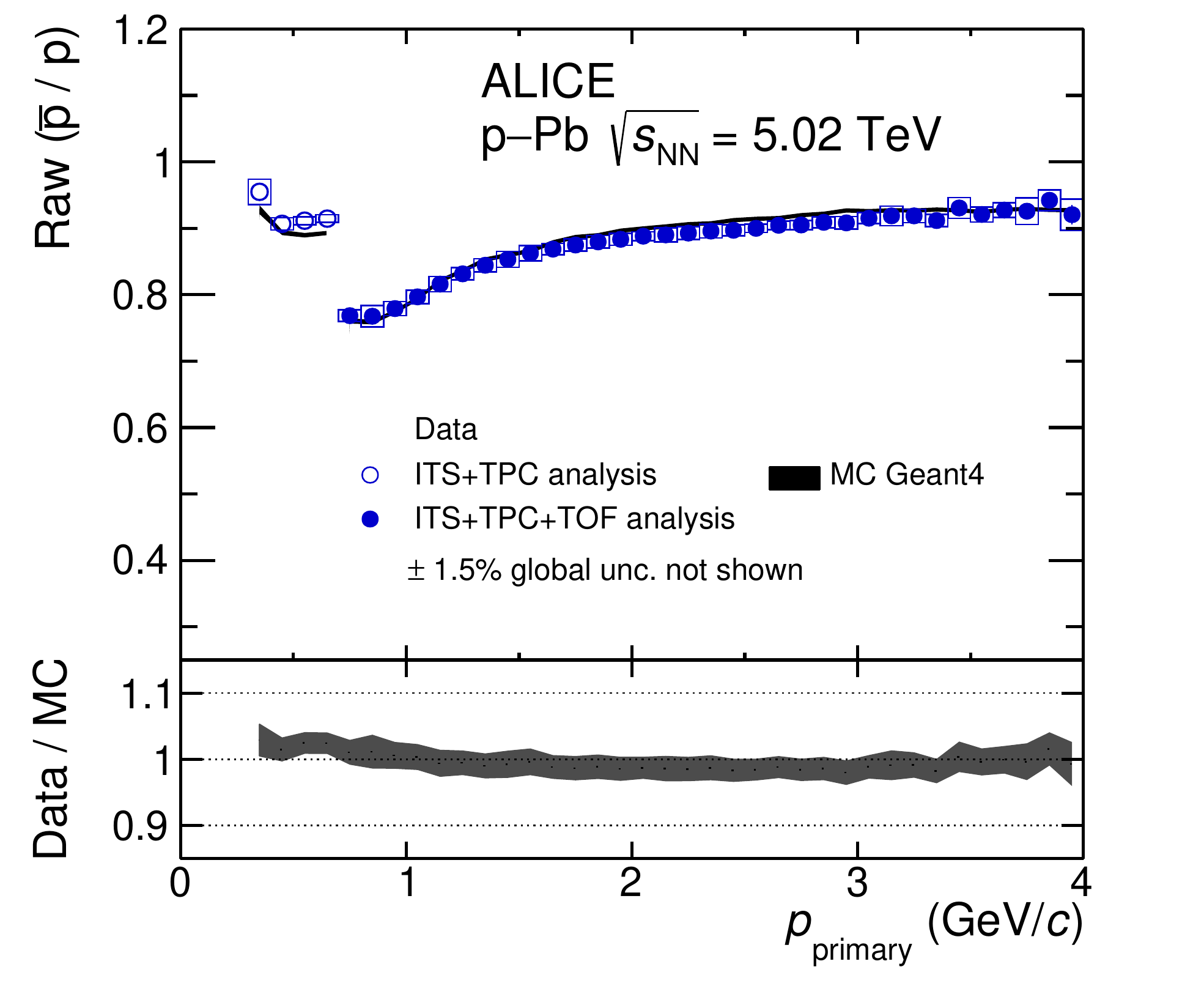}
\hfill
\includegraphics[width=0.49\linewidth]{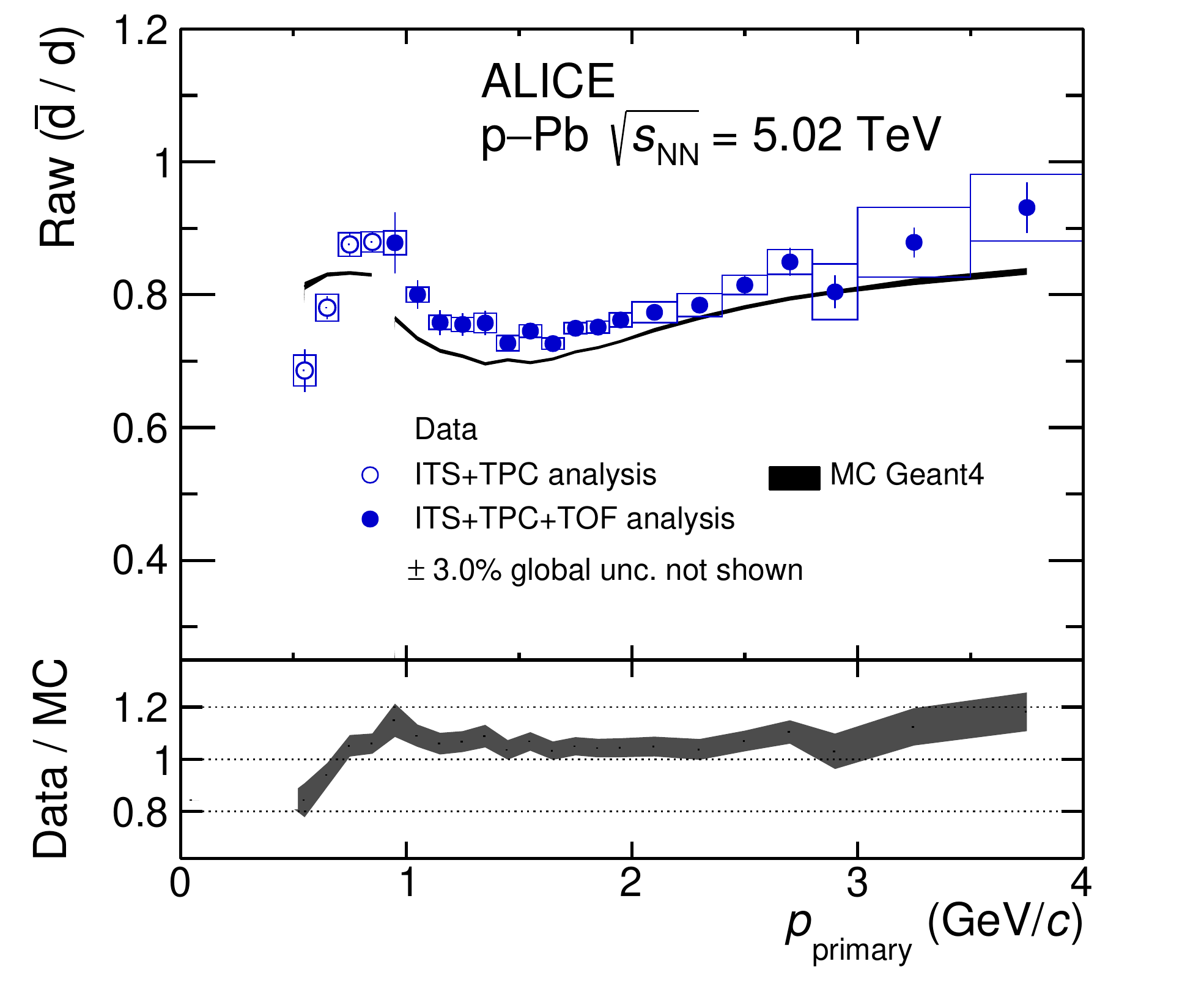}
\caption{Raw primary \RatioAntip (left) and \RatioAntid (right) ratios as a function of the momentum $p_{\rm primary}$. Experimental data are shown in blue, the statistical and systematic uncertainties are shown as vertical bars and boxes. The results from ALICE MC simulations based on \geantf using the FTFP\_INCLXX\_EMV physics list are shown in black. The width of the MC band represents the statistical uncertainty of the simulation. The global uncertainty due to the primordial ratio (1.5\% for $\bar{\rm p}/\rm{p}$ and 3\% for $\bar{\rm d}/\rm{d}$) is not shown in the top panels. The bottom panels display the ratios of experimental data to MC simulations with statistical, systematic and global uncertainties added in quadrature.}
\label{fig:RawRatios}
\end{figure}

The determination of the inelastic cross section requires precise knowledge of the ALICE detector material.

The MC parameterization of the ALICE material budget up to the outer TPC vessel was validated with photon conversion analyses within a precision of $\sim4.5\%$~\cite{Abelev:2014ffa} and it is shown in the supplemental material~\ref{SupplemtentalMaterialSection}. The ALICE detector material from the primary interaction point up to the TOF has an average atomic number of $\langle Z\rangle = 14.8$ and a mass number of $\langle A\rangle = 31.8$.
For the detector material up to the middle of the TPC, these values amount to $\langle Z\rangle = 8.5$ and $\langle A\rangle = 17.4$.
These values have been obtained by weighting the contribution from different materials with their density times length crossed by particles.
The selected (anti)proton and deuteron candidates include a substantial amount of background from secondary \mbox{(anti)}particles that originate from weak decays of hyperons or from spallation reactions in the detector material.
Following the procedure described in~\cite{Adam:2015vda,Abbas:2013rua,Aamodt:2010dx}, the contribution from secondary \mbox{(anti)}particles is subtracted by performing a fit to the distribution of the measured distance of closest approach (DCA) of track candidates to the primary vertex with templates from Monte Carlo (MC) simulations.
In contrast to secondary particles, primary particles point back to the primary vertex, hence a distinct structure peaked at zero in the DCA distribution characterizes the primary particles. Secondary particles correspond to a flat DCA distribution and their contribution can therefore be separated~\cite{Acharya:2017fvb,Adam:2015vda}.
The fraction of secondary \mbox{(anti)}protons is found to be around 20\% in the lowest momentum interval analysed ($0.3 \leq p_{\rm primary} < 0.4$ \GeVc) and decreases monotonically down to $\sim1.5\%$ at high momenta.
The main contribution of secondary \mbox{(anti)}protons stems from weak decays.
For deuterons, the dominant contribution of secondary particles comes from spallation processes in the detector material that lead to the ejection of fragments such as protons, neutrons or deuterons.
The fraction of secondary deuterons is found to be 23.5\% in the lowest momentum interval ($0.5 \leq p_{\mathrm{primary}} < 0.6$ \GeVc) and to decrease exponentially to negligible values at $p_{\mathrm{primary}} \sim 1.4$ \GeVc. 
For antiprotons and antideuterons the contribution from spallation processes is absent. The feed-down from weak decays of hyperons and hypernuclei has a negligible impact on the measured ratios~\cite{Abbas:2013rua,Acharya:2019rys,Adam:2015yta}.
Hence, the antideuteron sample is composed entirely from primaries. The total number of selected candidates amounts to $7.57\times10^{7}$ protons, $6.52\times 10^{7}$ antiprotons, $2.52\times 10^{5}$ deuterons and $1.98\times 10^{5}$ antideuterons. The momentum spectra are corrected for the background from secondary particles but not for the detector efficiency or losses of \mbox{(anti)}particles in the detector material, so they are referred to as raw primary spectra.
\begin{figure}[t]
\centering
\includegraphics[width=0.7\linewidth]{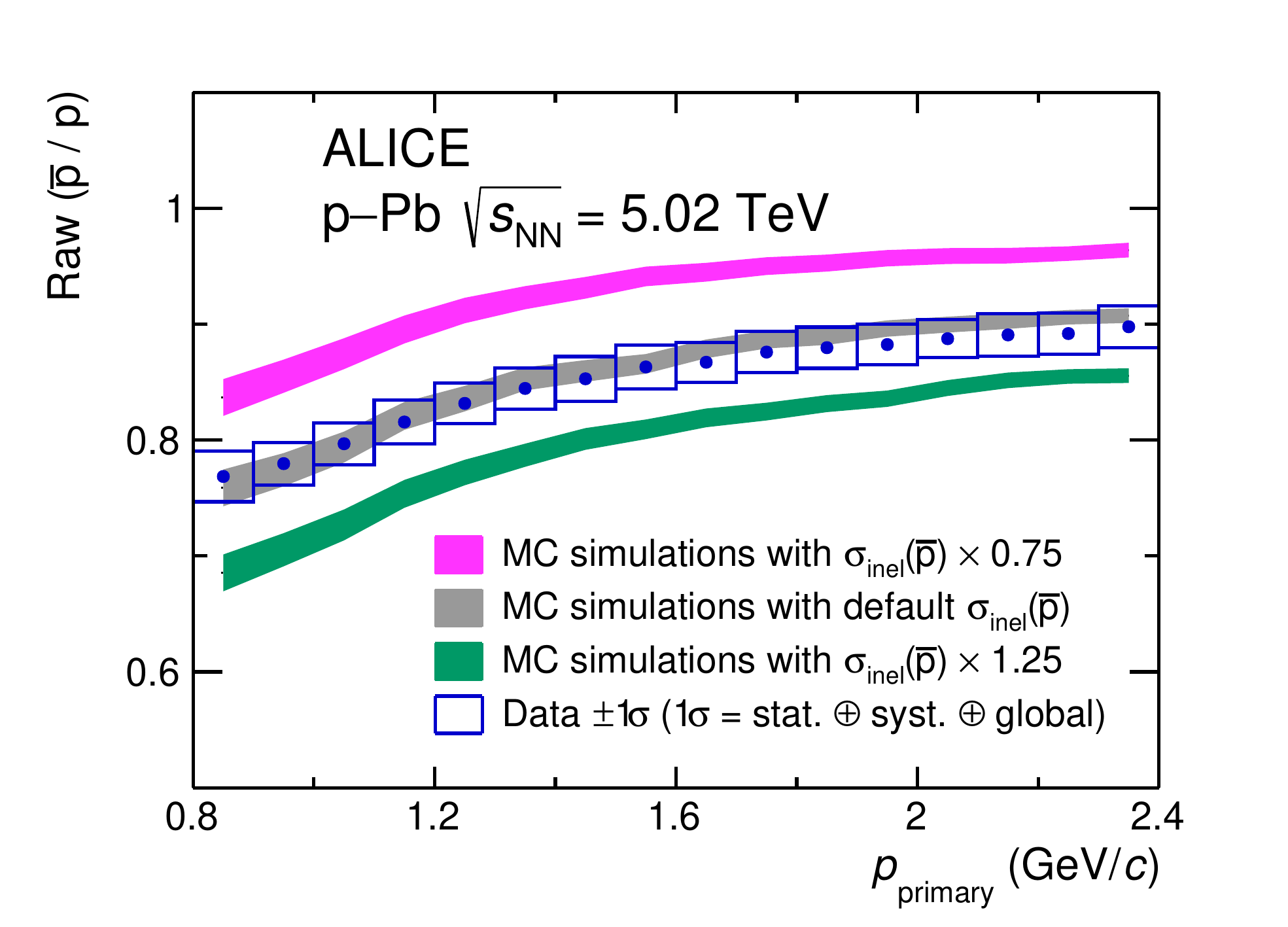}
\caption{Raw primary \RatioAntip ratio as a function of momentum. Blue boxes indicate $\pm1\sigma$ experimental limits. The results from MC simulations with varied \SigmaInelAntiP are shown as green and magenta bands, and gray band corresponds to the results with default \SigmaInelAntiP. The uncertainties on MC results include the variations of elastic cross sections and the variation of \SigmaInelP.
}
\label{fig:SigmaInelPVariation}
\end{figure}

Figure~\ref{fig:RawRatios} shows the \RatioAntip and \RatioAntid ratios as a function of $p_{\rm primary}$. 
The systematic uncertainties due to tracking, particle identification and contribution from secondaries are considered, and the total uncertainty is obtained as the quadratic sum of the individual contributions.
It increases from 1\% (2\%) at low momentum up to 2\% (6\%) in the high-momentum region for \RatioAntip (\RatioAntid).
The uncertainty on the primordial antimatter-to-matter ratio produced in collisions is considered as a global uncertainty.
The primordial \RatioAntip ratio $0.984 \pm 0.015$ is extrapolated from available measurements~\cite{Abbas:2013rua,Aamodt:2010dx} and, under the assumption that the \mbox{(anti)}deuteron yield is proportional to the squared yield of \mbox{(anti)}protons~\cite{Butler:1963pp,Scheibl:1998tk}, the primary \RatioAntid ratio amounts to $0.968 \pm 0.030$. These values are used as an input for detailed MC simulations based on the \geantf toolkit for the propagation of \mbox{(anti)}particles through the detector material~\cite{Agostinelli:2002hh}.
For the description of antinucleus--nucleus inelastic cross sections, \geantf relies on a Glauber calculation convoluted with a MC averaging method~\cite{Uzhinsky:2011zz}.
Figure~\ref{fig:RawRatios} shows that the \geantf-based simulations are able to describe the \RatioAntip ratio and are in qualitative agreement with the data for the \RatioAntid ratio.

The sensitivity of the antiparticle-to-particle ratios to the modifications of elastic and inelastic cross sections was benchmarked with the \RatioAntip measurement.
The \mbox{(anti)}proton cross sections have been measured by various experiments~\cite{Kuzichev:1993tn, Bianconi:2011zz, Schwaller:1979eu, Bauhoff:1986gcb, Bobchenko:1979hp, Nakamura:1984xw, Ashford:1985fh}, and the results are described well by the \geantf parameterization.
The blue boxes in Fig.~\ref{fig:SigmaInelPVariation} indicate the $\pm1\sigma$ limits for the measured \RatioAntip ratio, where $1\sigma$ corresponds to the quadratic sum of statistical, systematic and global uncertainties.
The green and magenta bands show the simulated ratios with a variation of $\pm25\%$ of the inelastic antiproton cross section along with the simulations using default cross section (gray band). Only a variation of the total inelastic cross section has been carried out.
The widths of the bands correspond to a quadratic sum of the contributions from two additional variations: i) the elastic cross sections of protons and antiprotons are changed independently by $\pm 20\%$, which leads to $\lesssim 1.5 \%$ modification of the ratio and ii) the inelastic proton--nucleus cross section is varied by $3.5\%$, which is the uncertainty of the \geantf parameterizations obtained from fits of the experimental data for this cross section. This variation yields a modification of about $0.5\%$ in the ratio. 
These systematic checks demonstrate that the antiparticle-to-particle ratio is mainly sensitive to the variation of the inelastic cross sections and can therefore be used to measure the antideuteron inelastic cross section.

\begin{figure}[hbt]
\begin{minipage}[h]{0.47\linewidth}
\center{\includegraphics[width=1\linewidth]{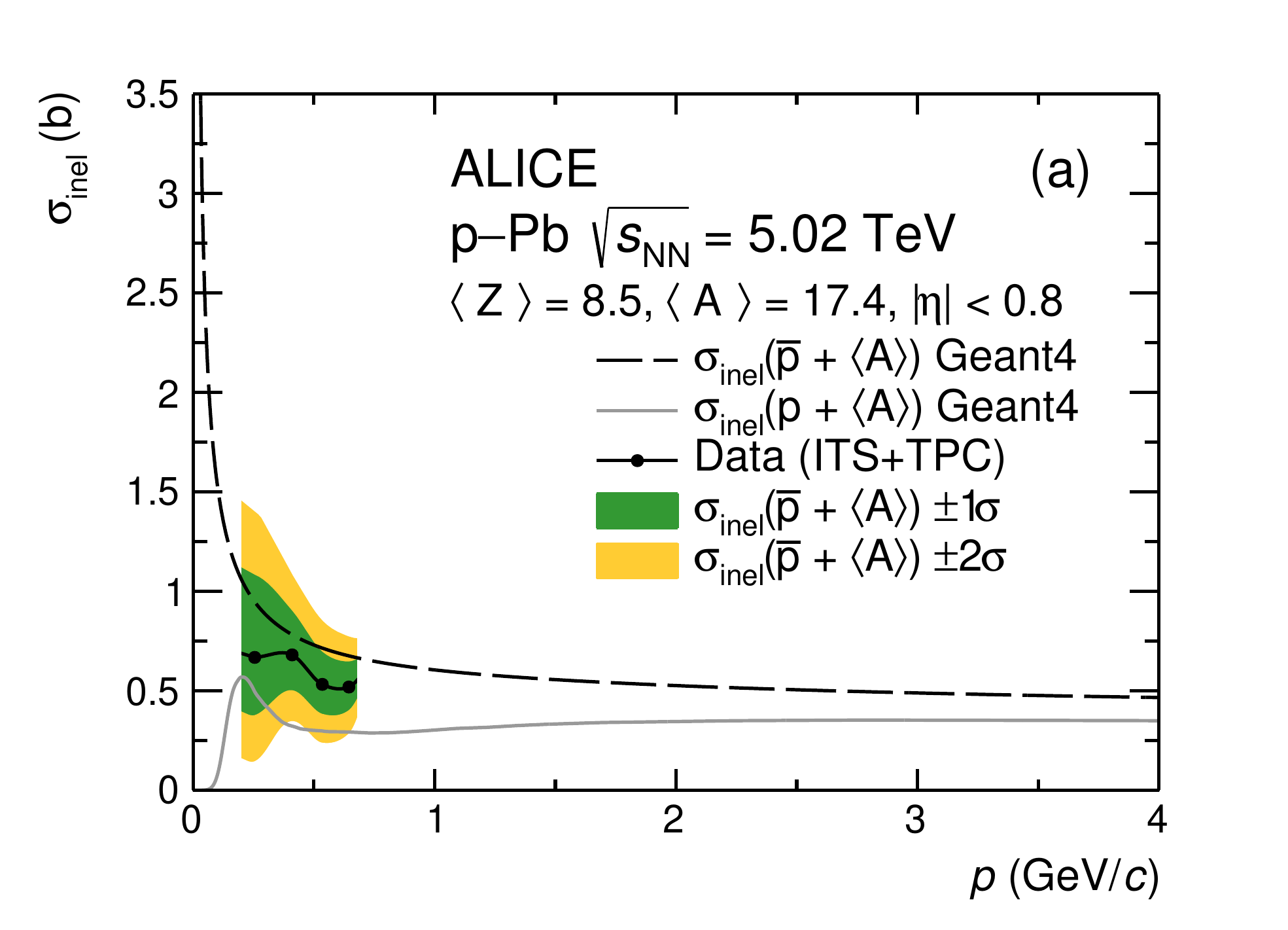}} 
\end{minipage}
\begin{minipage}[h]{0.47\linewidth}
\center{\includegraphics[width=1\linewidth]{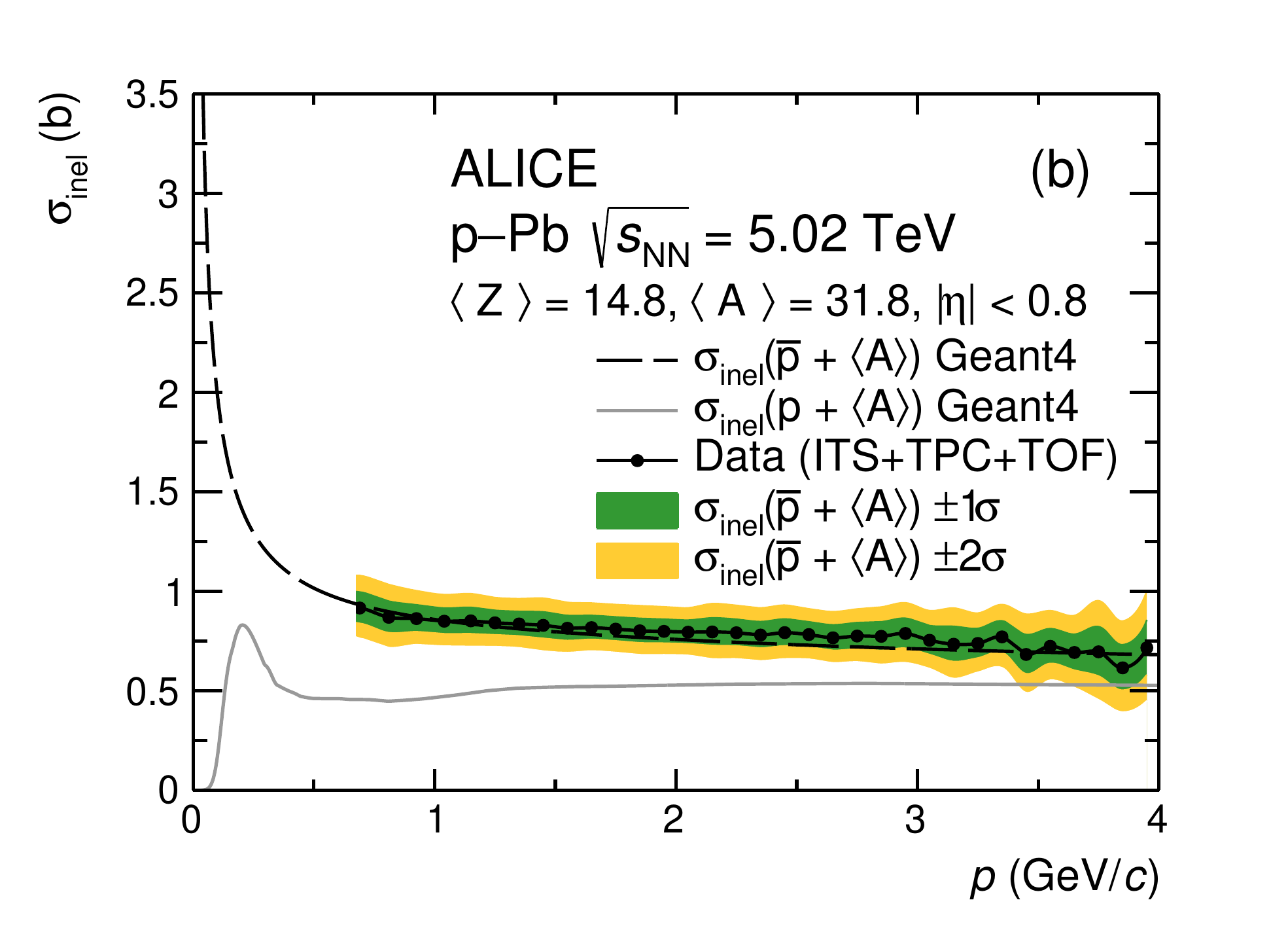}}
\end{minipage}
\begin{minipage}[h]{0.47\linewidth}
\center{\includegraphics[width=1\linewidth]{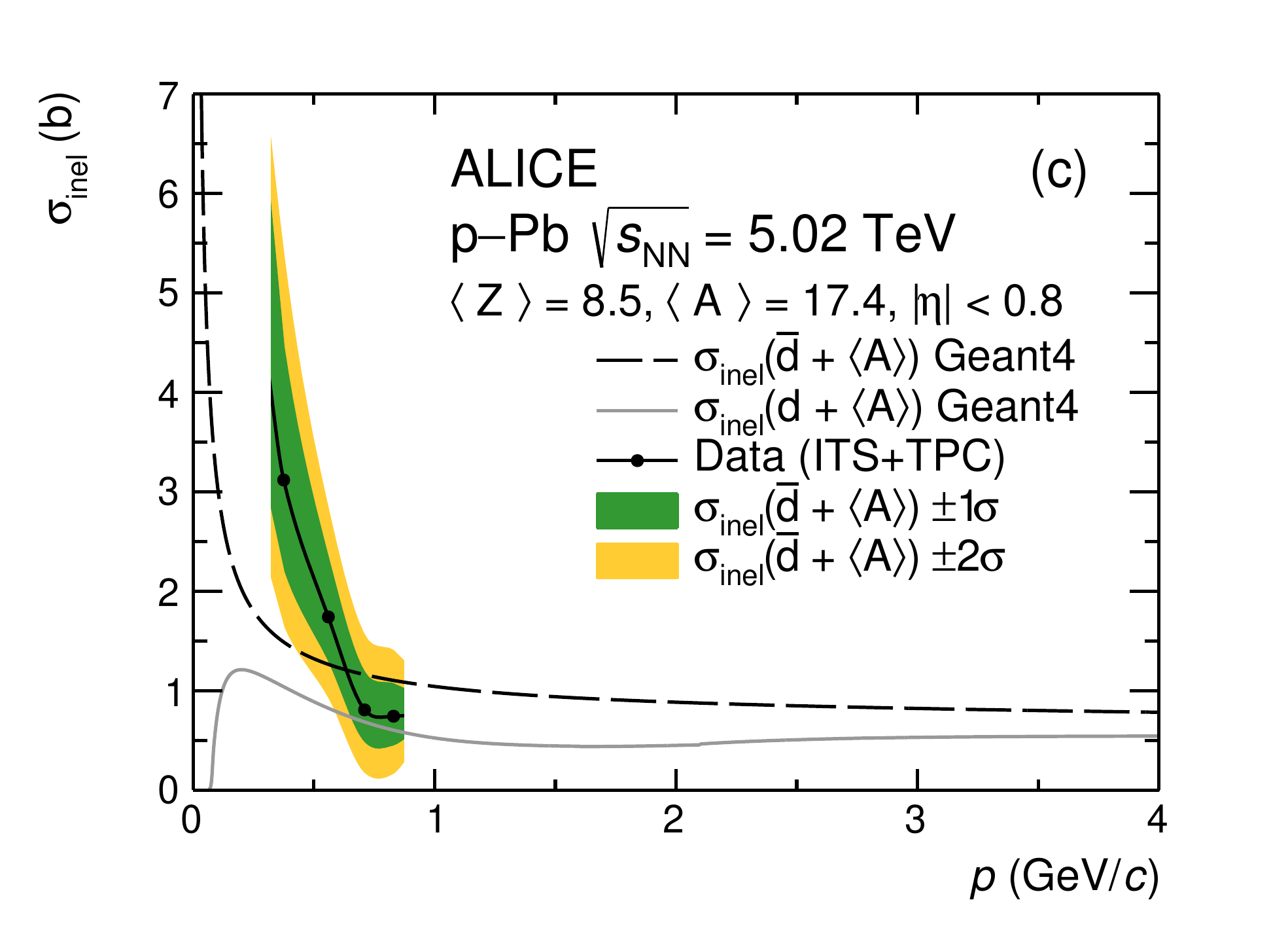}}
\end{minipage}
\hfill
\begin{minipage}[h]{0.47\linewidth}
\center{\includegraphics[width=1\linewidth]{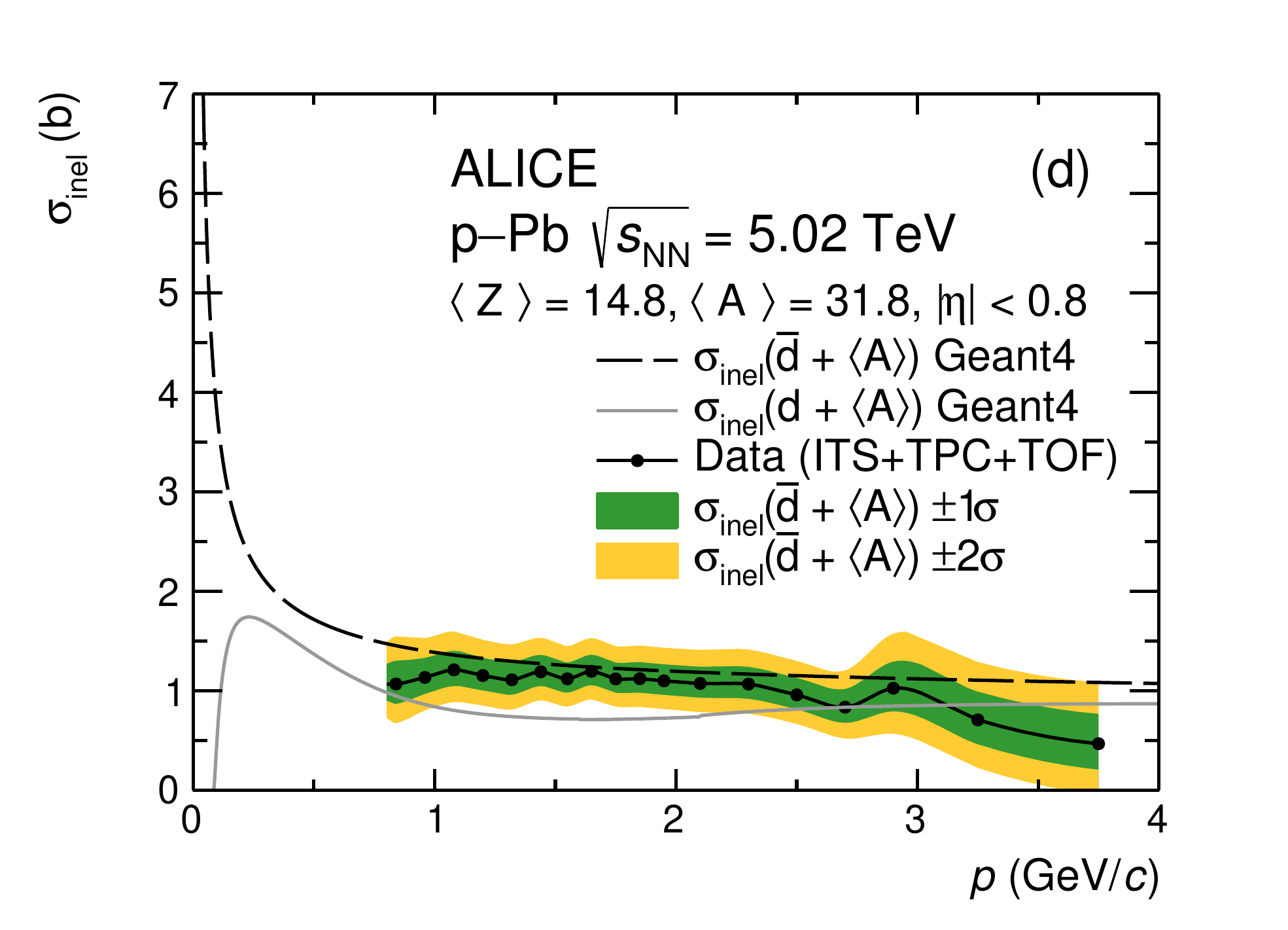}}
\end{minipage}
\caption{Inelastic interaction cross section for antiprotons and antideuterons on an average material element of the ALICE detector as a function of the momentum $p$ at which the interaction occurs. The top row shows the results for antiprotons, the bottom row for antideuterons and the results from the ITS+TPC (ITS+TPC+TOF) analysis are shown on the left (right). Dashed black lines represent the \geantf parameterizations for antinuclei, and full gray lines show the parameterizations for protons and deuterons. The experimental data points are shown connected by solid black lines, with green and orange bands corresponding to $\pm 1$ and $\pm 2\sigma$ constraints from the raw primary ratios.}
\label{fig:CrossSections_all}
\end{figure}

Extending this recipe, an iterative and momentum-dependent variation of \SigmaInelAntiP within the \geantf simulations was carried out to obtain \RatioAntip ratios that correspond to the $\pm1\sigma$ and $\pm2\sigma$ experimental limits.
The resulting $\pm1\sigma$ and $\pm2\sigma$ limits for \SigmaInelAntiP are presented in panels a) and b) of Fig.~\ref{fig:CrossSections_all} together with standard \geantf parameterizations.
Panel a) refers to the ITS+TPC analysis and hence corresponds to the inelastic interaction with nuclei that have average charge and mass number  $\langle Z\rangle = 8.5$ and $\langle A\rangle = 17.4$; panel b) refers to the analysis additionally employing the TOF and corresponds to $\langle Z\rangle = 14.8$ and $\langle A\rangle = 31.8$.
The inelastic cross sections shown in Fig.~\ref{fig:CrossSections_all} are estimated as a function of the momentum $p$ at which the inelastic interaction occurs. 
Due to the continuous energy loss of the particle inside the detector material, this momentum is lower than $p_{\rm primary}$ reconstructed at the primary vertex.

The corresponding correction is estimated using MC simulations by looking at the average values of the annihilation momentum distribution in each $p_{\rm primary}$ interval. The RMS of the distributions is then propagated to the uncertainty of the cross section measurement.
The minimum momentum reconstructed at the primary vertex amounts to $p_{\rm primary} = 0.3$~\GeVc for antiprotons and to $p_{\rm primary} = 0.5$~\GeVc for antideuterons, and the energy-loss correction transforms these values to $p = 0.18$~\GeVc and $p = 0.3$~\GeVc, correspondingly. For momenta $p > 0.7$~\GeVc, the antiproton inelastic cross section is found to be in good agreement with the \geantf parameterizations, which in turn describe well the existing experimental data~\cite{Uzhinsky:2011zz}.
Thus, these results validate the analysis procedure, which then can be applied to \mbox{(anti)}deuterons.

In contrast to antideuterons, the deuteron inelastic cross section was measured on several materials at various momenta~\cite{Auce:1996fe,Jaros:1977it}, and the data are well described by \geantf parameterizations.
The antideuteron inelastic cross section can therefore be constrained via the comparison of the experimental \RatioAntid ratio and the \geantf-based MC simulations with \SigmaInelAntiD varied in a similar way as for antiprotons.
For this purpose, the same uncertainties are considered: i) the variation of elastic cross sections of \mbox{(anti)}deuterons by $\pm 20\%$ that results in $\lesssim 2 \%$ deviation for the ratio, ii) the variation of the inelastic deuteron cross section by $7\%$ that corresponds to the precision of \geantf parameterizations ($\lesssim 1 \%$ uncertainty) and iii) the uncertainty from the primordial \RatioAntid ratio ($3.0\%$).

The resulting upper and lower limits on \SigmaInelAntiD 
for targets with $\langle Z\rangle = 8.5$, $\langle A\rangle = 17.4$ and $\langle Z\rangle = 14.8$, $\langle A\rangle = 31.8$ are shown in panels c) and d) of Fig.~\ref{fig:CrossSections_all}, respectively. 
The extracted inelastic cross sections presented here include all inelastic antideuteron processes where the antideuteron is destroyed and represent the first measurement in this low-momentum range.

While the measured \SigmaInelAntiD is found to be in agreement with the \geantf implementation within the $0.9 \leq p < 4.0$ \GeVc momentum range, it rises faster than the simulated parameterization in the momentum range $0.3 \leq p < 0.9$ \GeVc, reaching a maximal discrepancy of a factor $2.1$ in the interval of $0.3 \leq p < 0.47$ \GeVc.

These measurements can now help to better understand the antideuteron inelastic processes at low momenta and to improve the parameterization of the inelastic cross section used in \geantf.
Additionally, these results are now available for models of the propagation of antideuterons within the interstellar medium~\cite{Duperray:2005si, Donato:2008yx,Kachelriess:2020uoh} and will impact the flux expectations at low momentum near Earth.

In summary, we have shown how the ALICE detector can be used as an absorber to study the antinuclei inelastic scattering cross section on detector material. The antiparticle-to-particle ratios method was validated using (anti)protons and the sensitivity of the ratio to the variation of the inelastic cross section was demonstrated. In this way, the first measurement of the inelastic scattering cross section of antideuterons was performed on an effective target with mean charge number $\langle Z\rangle = 8.5$ and mass number $\langle A\rangle = 17.4$ in the momentum range $0.3 \leq p < 0.9$ \GeVc, and with $\langle Z\rangle = 14.8$ and $\langle A\rangle = 31.8$ in $0.9 \leq p < 4.0$ \GeVc.
These cross sections can now be used in propagation models of antideuterons within the interstellar medium for dark-matter searches.
Future studies of high-statistics \pp, \pPb and \PbPb data collected during the second (2015--2018) and third (scheduled to start in 2021) LHC run campaigns should allow the measurement of inelastic cross sections of heavier antinuclei such as \AntiHeThree and \AntiHeFour in a similar way and the improvement of the current antideuteron results.


\newenvironment{acknowledgement}{\relax}{\relax}
\begin{acknowledgement}
\section*{Acknowledgements}

The ALICE Collaboration would like to thank all its engineers and technicians for their invaluable contributions to the construction of the experiment and the CERN accelerator teams for the outstanding performance of the LHC complex.
The ALICE Collaboration gratefully acknowledges the resources and support provided by all Grid centres and the Worldwide LHC Computing Grid (WLCG) collaboration.
The ALICE Collaboration acknowledges the following funding agencies for their support in building and running the ALICE detector:
A. I. Alikhanyan National Science Laboratory (Yerevan Physics Institute) Foundation (ANSL), State Committee of Science and World Federation of Scientists (WFS), Armenia;
Austrian Academy of Sciences, Austrian Science Fund (FWF): [M 2467-N36] and Nationalstiftung f\"{u}r Forschung, Technologie und Entwicklung, Austria;
Ministry of Communications and High Technologies, National Nuclear Research Center, Azerbaijan;
Conselho Nacional de Desenvolvimento Cient\'{\i}fico e Tecnol\'{o}gico (CNPq), Financiadora de Estudos e Projetos (Finep), Funda\c{c}\~{a}o de Amparo \`{a} Pesquisa do Estado de S\~{a}o Paulo (FAPESP) and Universidade Federal do Rio Grande do Sul (UFRGS), Brazil;
Ministry of Education of China (MOEC) , Ministry of Science \& Technology of China (MSTC) and National Natural Science Foundation of China (NSFC), China;
Ministry of Science and Education and Croatian Science Foundation, Croatia;
Centro de Aplicaciones Tecnol\'{o}gicas y Desarrollo Nuclear (CEADEN), Cubaenerg\'{\i}a, Cuba;
Ministry of Education, Youth and Sports of the Czech Republic, Czech Republic;
The Danish Council for Independent Research | Natural Sciences, the VILLUM FONDEN and Danish National Research Foundation (DNRF), Denmark;
Helsinki Institute of Physics (HIP), Finland;
Commissariat \`{a} l'Energie Atomique (CEA) and Institut National de Physique Nucl\'{e}aire et de Physique des Particules (IN2P3) and Centre National de la Recherche Scientifique (CNRS), France;
Bundesministerium f\"{u}r Bildung und Forschung (BMBF) and GSI Helmholtzzentrum f\"{u}r Schwerionenforschung GmbH, Germany;
General Secretariat for Research and Technology, Ministry of Education, Research and Religions, Greece;
National Research, Development and Innovation Office, Hungary;
Department of Atomic Energy Government of India (DAE), Department of Science and Technology, Government of India (DST), University Grants Commission, Government of India (UGC) and Council of Scientific and Industrial Research (CSIR), India;
Indonesian Institute of Science, Indonesia;
Centro Fermi - Museo Storico della Fisica e Centro Studi e Ricerche Enrico Fermi and Istituto Nazionale di Fisica Nucleare (INFN), Italy;
Institute for Innovative Science and Technology , Nagasaki Institute of Applied Science (IIST), Japanese Ministry of Education, Culture, Sports, Science and Technology (MEXT) and Japan Society for the Promotion of Science (JSPS) KAKENHI, Japan;
Consejo Nacional de Ciencia (CONACYT) y Tecnolog\'{i}a, through Fondo de Cooperaci\'{o}n Internacional en Ciencia y Tecnolog\'{i}a (FONCICYT) and Direcci\'{o}n General de Asuntos del Personal Academico (DGAPA), Mexico;
Nederlandse Organisatie voor Wetenschappelijk Onderzoek (NWO), Netherlands;
The Research Council of Norway, Norway;
Commission on Science and Technology for Sustainable Development in the South (COMSATS), Pakistan;
Pontificia Universidad Cat\'{o}lica del Per\'{u}, Peru;
Ministry of Science and Higher Education, National Science Centre and WUT ID-UB, Poland;
Korea Institute of Science and Technology Information and National Research Foundation of Korea (NRF), Republic of Korea;
Ministry of Education and Scientific Research, Institute of Atomic Physics and Ministry of Research and Innovation and Institute of Atomic Physics, Romania;
Joint Institute for Nuclear Research (JINR), Ministry of Education and Science of the Russian Federation, National Research Centre Kurchatov Institute, Russian Science Foundation and Russian Foundation for Basic Research, Russia;
Ministry of Education, Science, Research and Sport of the Slovak Republic, Slovakia;
National Research Foundation of South Africa, South Africa;
Swedish Research Council (VR) and Knut \& Alice Wallenberg Foundation (KAW), Sweden;
European Organization for Nuclear Research, Switzerland;
Suranaree University of Technology (SUT), National Science and Technology Development Agency (NSDTA) and Office of the Higher Education Commission under NRU project of Thailand, Thailand;
Turkish Atomic Energy Agency (TAEK), Turkey;
National Academy of  Sciences of Ukraine, Ukraine;
Science and Technology Facilities Council (STFC), United Kingdom;
National Science Foundation of the United States of America (NSF) and United States Department of Energy, Office of Nuclear Physics (DOE NP), United States of America.
\end{acknowledgement}

\bibliographystyle{utphys}   
\bibliography{bibliography}

\newpage
\appendix

%
%

\section{The ALICE Collaboration}
\label{app:collab}

\begingroup
\small
\begin{flushleft}
S.~Acharya\Irefn{org141}\And 
D.~Adamov\'{a}\Irefn{org95}\And 
A.~Adler\Irefn{org74}\And 
J.~Adolfsson\Irefn{org81}\And 
M.M.~Aggarwal\Irefn{org100}\And 
G.~Aglieri Rinella\Irefn{org34}\And 
M.~Agnello\Irefn{org30}\And 
N.~Agrawal\Irefn{org10}\textsuperscript{,}\Irefn{org54}\And 
Z.~Ahammed\Irefn{org141}\And 
S.~Ahmad\Irefn{org16}
% ...
\And
\renewcommand\labelenumi{\textsuperscript{\theenumi}~}

\section*{Affiliation notes}
\renewcommand\theenumi{\roman{enumi}}
\begin{Authlist}
\item \Adef{org*}Deceased
\item \Adef{orgI}Italian National Agency for New Technologies, Energy and Sustainable Economic Development (ENEA), Bologna, Italy
\item \Adef{orgII}Dipartimento DET del Politecnico di Torino, Turin, Italy
\item \Adef{orgIII}M.V. Lomonosov Moscow State University, D.V. Skobeltsyn Institute of Nuclear, Physics, Moscow, Russia
\item \Adef{orgIV}Department of Applied Physics, Aligarh Muslim University, Aligarh, India
\item \Adef{orgV}Institute of Theoretical Physics, University of Wroclaw, Poland
\end{Authlist}

\section*{Collaboration Institutes}
\renewcommand\theenumi{\arabic{enumi}~}
\begin{Authlist}
\item \Idef{org1}A.I. Alikhanyan National Science Laboratory (Yerevan Physics Institute) Foundation, Yerevan, Armenia
\item \Idef{org2}Bogolyubov Institute for Theoretical Physics, National Academy of Sciences of Ukraine, Kiev, Ukraine
\item \Idef{org3}Bose Institute, Department of Physics  and Centre for Astroparticle Physics and Space Science (CAPSS), Kolkata, India
\item \Idef{org4}Budker Institute for Nuclear Physics, Novosibirsk, Russia
\item \Idef{org5}California Polytechnic State University, San Luis Obispo, California, United States
\item \Idef{org6}Central China Normal University, Wuhan, China
\item \Idef{org7}Centre de Calcul de l'IN2P3, Villeurbanne, Lyon, France
\item \Idef{org8}Centro de Aplicaciones Tecnol\'{o}gicas y Desarrollo Nuclear (CEADEN), Havana, Cuba
\item \Idef{org9}Centro de Investigaci\'{o}n y de Estudios Avanzados (CINVESTAV), Mexico City and M\'{e}rida, Mexico
\item \Idef{org10}Centro Fermi - Museo Storico della Fisica e Centro Studi e Ricerche ``Enrico Fermi', Rome, Italy
\item \Idef{org11}Chicago State University, Chicago, Illinois, United States
\item \Idef{org12}China Institute of Atomic Energy, Beijing, China
\item \Idef{org13}Comenius University Bratislava, Faculty of Mathematics, Physics and Informatics, Bratislava, Slovakia
\item \Idef{org14}COMSATS University Islamabad, Islamabad, Pakistan
\item \Idef{org15}Creighton University, Omaha, Nebraska, United States
\item \Idef{org16}Department of Physics, Aligarh Muslim University, Aligarh, India
\item \Idef{org17}Department of Physics, Pusan National University, Pusan, Republic of Korea
\item \Idef{org18}Department of Physics, Sejong University, Seoul, Republic of Korea
\item \Idef{org19}Department of Physics, University of California, Berkeley, California, United States
\item \Idef{org20}Department of Physics, University of Oslo, Oslo, Norway
\item \Idef{org21}Department of Physics and Technology, University of Bergen, Bergen, Norway
\item \Idef{org22}Dipartimento di Fisica dell'Universit\`{a} 'La Sapienza' and Sezione INFN, Rome, Italy
\item \Idef{org23}Dipartimento di Fisica dell'Universit\`{a} and Sezione INFN, Cagliari, Italy
\item \Idef{org24}Dipartimento di Fisica dell'Universit\`{a} and Sezione INFN, Trieste, Italy
\item \Idef{org25}Dipartimento di Fisica dell'Universit\`{a} and Sezione INFN, Turin, Italy
\item \Idef{org26}Dipartimento di Fisica e Astronomia dell'Universit\`{a} and Sezione INFN, Bologna, Italy
\item \Idef{org27}Dipartimento di Fisica e Astronomia dell'Universit\`{a} and Sezione INFN, Catania, Italy
\item \Idef{org28}Dipartimento di Fisica e Astronomia dell'Universit\`{a} and Sezione INFN, Padova, Italy
\item \Idef{org29}Dipartimento di Fisica `E.R.~Caianiello' dell'Universit\`{a} and Gruppo Collegato INFN, Salerno, Italy
\item \Idef{org30}Dipartimento DISAT del Politecnico and Sezione INFN, Turin, Italy
\item \Idef{org31}Dipartimento di Scienze e Innovazione Tecnologica dell'Universit\`{a} del Piemonte Orientale and INFN Sezione di Torino, Alessandria, Italy
\item \Idef{org32}Dipartimento di Scienze MIFT, Universit\`{a} di Messina, Messina, Italy
\item \Idef{org33}Dipartimento Interateneo di Fisica `M.~Merlin' and Sezione INFN, Bari, Italy
\item \Idef{org34}European Organization for Nuclear Research (CERN), Geneva, Switzerland
\item \Idef{org35}Faculty of Electrical Engineering, Mechanical Engineering and Naval Architecture, University of Split, Split, Croatia
\item \Idef{org36}Faculty of Engineering and Science, Western Norway University of Applied Sciences, Bergen, Norway
\item \Idef{org37}Faculty of Nuclear Sciences and Physical Engineering, Czech Technical University in Prague, Prague, Czech Republic
\item \Idef{org38}Faculty of Science, P.J.~\v{S}af\'{a}rik University, Ko\v{s}ice, Slovakia
\item \Idef{org39}Frankfurt Institute for Advanced Studies, Johann Wolfgang Goethe-Universit\"{a}t Frankfurt, Frankfurt, Germany
\item \Idef{org40}Fudan University, Shanghai, China
\item \Idef{org41}Gangneung-Wonju National University, Gangneung, Republic of Korea
\item \Idef{org42}Gauhati University, Department of Physics, Guwahati, India
\item \Idef{org43}Helmholtz-Institut f\"{u}r Strahlen- und Kernphysik, Rheinische Friedrich-Wilhelms-Universit\"{a}t Bonn, Bonn, Germany
\item \Idef{org44}Helsinki Institute of Physics (HIP), Helsinki, Finland
\item \Idef{org45}High Energy Physics Group,  Universidad Aut\'{o}noma de Puebla, Puebla, Mexico
\item \Idef{org46}Hiroshima University, Hiroshima, Japan
\item \Idef{org47}Hochschule Worms, Zentrum  f\"{u}r Technologietransfer und Telekommunikation (ZTT), Worms, Germany
\item \Idef{org48}Horia Hulubei National Institute of Physics and Nuclear Engineering, Bucharest, Romania
\item \Idef{org49}Indian Institute of Technology Bombay (IIT), Mumbai, India
\item \Idef{org50}Indian Institute of Technology Indore, Indore, India
\item \Idef{org51}Indonesian Institute of Sciences, Jakarta, Indonesia
\item \Idef{org52}INFN, Laboratori Nazionali di Frascati, Frascati, Italy
\item \Idef{org53}INFN, Sezione di Bari, Bari, Italy
\item \Idef{org54}INFN, Sezione di Bologna, Bologna, Italy
\item \Idef{org55}INFN, Sezione di Cagliari, Cagliari, Italy
\item \Idef{org56}INFN, Sezione di Catania, Catania, Italy
\item \Idef{org57}INFN, Sezione di Padova, Padova, Italy
\item \Idef{org58}INFN, Sezione di Roma, Rome, Italy
\item \Idef{org59}INFN, Sezione di Torino, Turin, Italy
\item \Idef{org60}INFN, Sezione di Trieste, Trieste, Italy
\item \Idef{org61}Inha University, Incheon, Republic of Korea
\item \Idef{org62}Institute for Nuclear Research, Academy of Sciences, Moscow, Russia
\item \Idef{org63}Institute for Subatomic Physics, Utrecht University/Nikhef, Utrecht, Netherlands
\item \Idef{org64}Institute of Experimental Physics, Slovak Academy of Sciences, Ko\v{s}ice, Slovakia
\item \Idef{org65}Institute of Physics, Homi Bhabha National Institute, Bhubaneswar, India
\item \Idef{org66}Institute of Physics of the Czech Academy of Sciences, Prague, Czech Republic
\item \Idef{org67}Institute of Space Science (ISS), Bucharest, Romania
\item \Idef{org68}Institut f\"{u}r Kernphysik, Johann Wolfgang Goethe-Universit\"{a}t Frankfurt, Frankfurt, Germany
\item \Idef{org69}Instituto de Ciencias Nucleares, Universidad Nacional Aut\'{o}noma de M\'{e}xico, Mexico City, Mexico
\item \Idef{org70}Instituto de F\'{i}sica, Universidade Federal do Rio Grande do Sul (UFRGS), Porto Alegre, Brazil
\item \Idef{org71}Instituto de F\'{\i}sica, Universidad Nacional Aut\'{o}noma de M\'{e}xico, Mexico City, Mexico
\item \Idef{org72}iThemba LABS, National Research Foundation, Somerset West, South Africa
\item \Idef{org73}Jeonbuk National University, Jeonju, Republic of Korea
\item \Idef{org74}Johann-Wolfgang-Goethe Universit\"{a}t Frankfurt Institut f\"{u}r Informatik, Fachbereich Informatik und Mathematik, Frankfurt, Germany
\item \Idef{org75}Joint Institute for Nuclear Research (JINR), Dubna, Russia
\item \Idef{org76}Korea Institute of Science and Technology Information, Daejeon, Republic of Korea
\item \Idef{org77}KTO Karatay University, Konya, Turkey
\item \Idef{org78}Laboratoire de Physique des 2 Infinis, Irène Joliot-Curie, Orsay, France
\item \Idef{org79}Laboratoire de Physique Subatomique et de Cosmologie, Universit\'{e} Grenoble-Alpes, CNRS-IN2P3, Grenoble, France
\item \Idef{org80}Lawrence Berkeley National Laboratory, Berkeley, California, United States
\item \Idef{org81}Lund University Department of Physics, Division of Particle Physics, Lund, Sweden
\item \Idef{org82}Nagasaki Institute of Applied Science, Nagasaki, Japan
\item \Idef{org83}Nara Women{'}s University (NWU), Nara, Japan
\item \Idef{org84}National and Kapodistrian University of Athens, School of Science, Department of Physics , Athens, Greece
\item \Idef{org85}National Centre for Nuclear Research, Warsaw, Poland
\item \Idef{org86}National Institute of Science Education and Research, Homi Bhabha National Institute, Jatni, India
\item \Idef{org87}National Nuclear Research Center, Baku, Azerbaijan
\item \Idef{org88}National Research Centre Kurchatov Institute, Moscow, Russia
\item \Idef{org89}Niels Bohr Institute, University of Copenhagen, Copenhagen, Denmark
\item \Idef{org90}Nikhef, National institute for subatomic physics, Amsterdam, Netherlands
\item \Idef{org91}NRC Kurchatov Institute IHEP, Protvino, Russia
\item \Idef{org92}NRC «Kurchatov Institute»  - ITEP, Moscow, Russia
\item \Idef{org93}NRNU Moscow Engineering Physics Institute, Moscow, Russia
\item \Idef{org94}Nuclear Physics Group, STFC Daresbury Laboratory, Daresbury, United Kingdom
\item \Idef{org95}Nuclear Physics Institute of the Czech Academy of Sciences, \v{R}e\v{z} u Prahy, Czech Republic
\item \Idef{org96}Oak Ridge National Laboratory, Oak Ridge, Tennessee, United States
\item \Idef{org97}Ohio State University, Columbus, Ohio, United States
\item \Idef{org98}Petersburg Nuclear Physics Institute, Gatchina, Russia
\item \Idef{org99}Physics department, Faculty of science, University of Zagreb, Zagreb, Croatia
\item \Idef{org100}Physics Department, Panjab University, Chandigarh, India
\item \Idef{org101}Physics Department, University of Jammu, Jammu, India
\item \Idef{org102}Physics Department, University of Rajasthan, Jaipur, India
\item \Idef{org103}Physikalisches Institut, Eberhard-Karls-Universit\"{a}t T\"{u}bingen, T\"{u}bingen, Germany
\item \Idef{org104}Physikalisches Institut, Ruprecht-Karls-Universit\"{a}t Heidelberg, Heidelberg, Germany
\item \Idef{org105}Physik Department, Technische Universit\"{a}t M\"{u}nchen, Munich, Germany
\item \Idef{org106}Politecnico di Bari, Bari, Italy
\item \Idef{org107}Research Division and ExtreMe Matter Institute EMMI, GSI Helmholtzzentrum f\"ur Schwerionenforschung GmbH, Darmstadt, Germany
\item \Idef{org108}Rudjer Bo\v{s}kovi\'{c} Institute, Zagreb, Croatia
\item \Idef{org109}Russian Federal Nuclear Center (VNIIEF), Sarov, Russia
\item \Idef{org110}Saha Institute of Nuclear Physics, Homi Bhabha National Institute, Kolkata, India
\item \Idef{org111}School of Physics and Astronomy, University of Birmingham, Birmingham, United Kingdom
\item \Idef{org112}Secci\'{o}n F\'{\i}sica, Departamento de Ciencias, Pontificia Universidad Cat\'{o}lica del Per\'{u}, Lima, Peru
\item \Idef{org113}St. Petersburg State University, St. Petersburg, Russia
\item \Idef{org114}Stefan Meyer Institut f\"{u}r Subatomare Physik (SMI), Vienna, Austria
\item \Idef{org115}SUBATECH, IMT Atlantique, Universit\'{e} de Nantes, CNRS-IN2P3, Nantes, France
\item \Idef{org116}Suranaree University of Technology, Nakhon Ratchasima, Thailand
\item \Idef{org117}Technical University of Ko\v{s}ice, Ko\v{s}ice, Slovakia
\item \Idef{org118}The Henryk Niewodniczanski Institute of Nuclear Physics, Polish Academy of Sciences, Cracow, Poland
\item \Idef{org119}The University of Texas at Austin, Austin, Texas, United States
\item \Idef{org120}Universidad Aut\'{o}noma de Sinaloa, Culiac\'{a}n, Mexico
\item \Idef{org121}Universidade de S\~{a}o Paulo (USP), S\~{a}o Paulo, Brazil
\item \Idef{org122}Universidade Estadual de Campinas (UNICAMP), Campinas, Brazil
\item \Idef{org123}Universidade Federal do ABC, Santo Andre, Brazil
\item \Idef{org124}University of Cape Town, Cape Town, South Africa
\item \Idef{org125}University of Houston, Houston, Texas, United States
\item \Idef{org126}University of Jyv\"{a}skyl\"{a}, Jyv\"{a}skyl\"{a}, Finland
\item \Idef{org127}University of Liverpool, Liverpool, United Kingdom
\item \Idef{org128}University of Science and Technology of China, Hefei, China
\item \Idef{org129}University of South-Eastern Norway, Tonsberg, Norway
\item \Idef{org130}University of Tennessee, Knoxville, Tennessee, United States
\item \Idef{org131}University of the Witwatersrand, Johannesburg, South Africa
\item \Idef{org132}University of Tokyo, Tokyo, Japan
\item \Idef{org133}University of Tsukuba, Tsukuba, Japan
\item \Idef{org134}Universit\'{e} Clermont Auvergne, CNRS/IN2P3, LPC, Clermont-Ferrand, France
\item \Idef{org135}Universit\'{e} de Lyon, Universit\'{e} Lyon 1, CNRS/IN2P3, IPN-Lyon, Villeurbanne, Lyon, France
\item \Idef{org136}Universit\'{e} de Strasbourg, CNRS, IPHC UMR 7178, F-67000 Strasbourg, France, Strasbourg, France
\item \Idef{org137}Universit\'{e} Paris-Saclay Centre d'Etudes de Saclay (CEA), IRFU, D\'{e}partment de Physique Nucl\'{e}aire (DPhN), Saclay, France
\item \Idef{org138}Universit\`{a} degli Studi di Foggia, Foggia, Italy
\item \Idef{org139}Universit\`{a} degli Studi di Pavia, Pavia, Italy
\item \Idef{org140}Universit\`{a} di Brescia, Brescia, Italy
\item \Idef{org141}Variable Energy Cyclotron Centre, Homi Bhabha National Institute, Kolkata, India
\item \Idef{org142}Warsaw University of Technology, Warsaw, Poland
\item \Idef{org143}Wayne State University, Detroit, Michigan, United States
\item \Idef{org144}Westf\"{a}lische Wilhelms-Universit\"{a}t M\"{u}nster, Institut f\"{u}r Kernphysik, M\"{u}nster, Germany
\item \Idef{org145}Wigner Research Centre for Physics, Budapest, Hungary
\item \Idef{org146}Yale University, New Haven, Connecticut, United States
\item \Idef{org147}Yonsei University, Seoul, Republic of Korea
\end{Authlist}
\endgroup

\newpage

\section{Supplemental material}\label{SupplemtentalMaterialSection}

\begin{figure}[!ht]
    \centering
    \includegraphics[width=\textwidth]{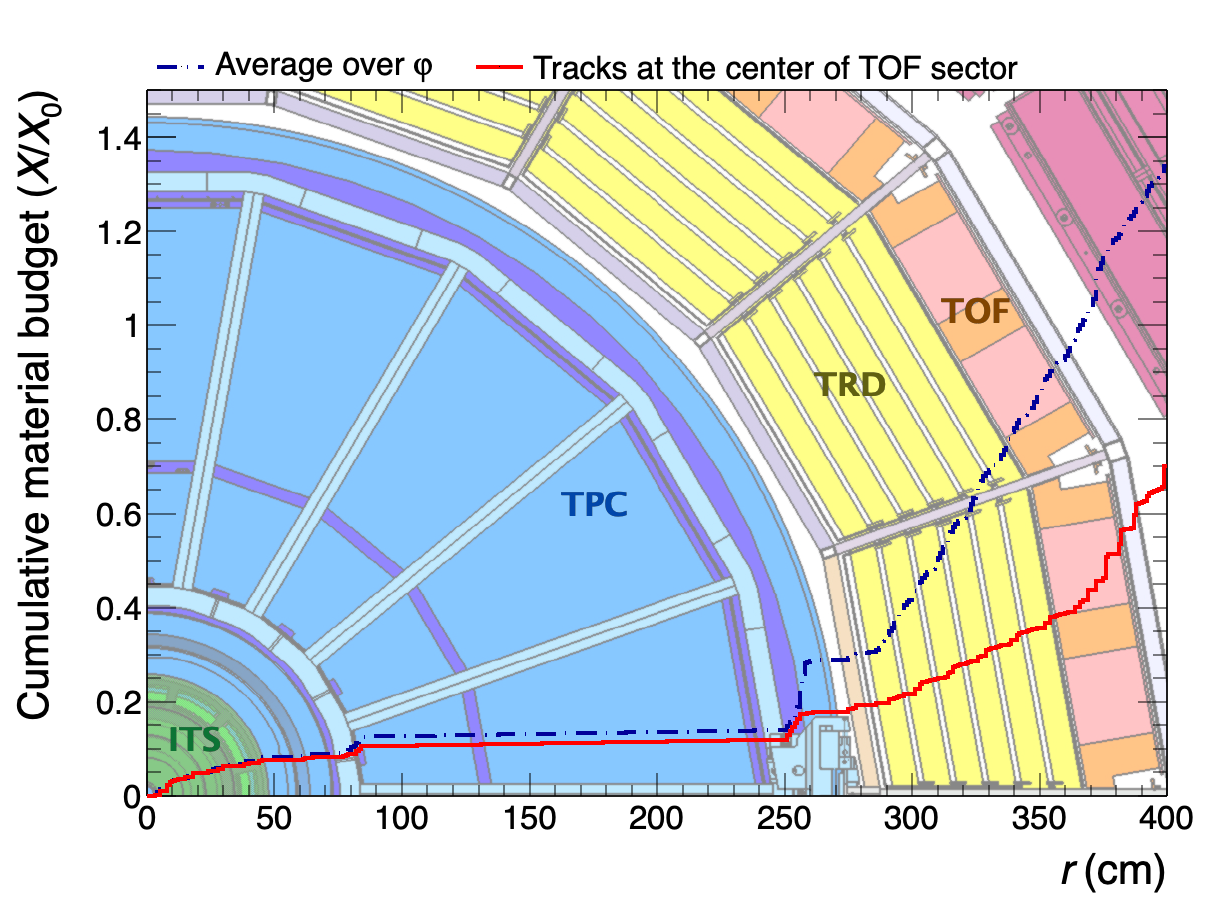}
    \caption{Cumulative distribution of the material in the ALICE apparatus as a function of the radial distance from the beam line. The results are shown for straight primary tracks emitted perpendicularly to the beam line either at the center of the TOF sectors (red line) or averaged over azimuth (blue line). The cross section on the beam-transverse plane of the different detector parts at the end cap is depicted with different colours in the background.}
    \label{fig:material-sketch}
\end{figure}
\end{document}